\documentclass[preprint,aps,nofootinbib]{revtex4}
\usepackage{graphicx}
\usepackage{epstopdf}
\usepackage{subfigure}
\usepackage{textcomp}
\usepackage{dsfont}
\usepackage{amsfonts}
\usepackage{amsmath}
\usepackage{relsize}
\usepackage{amssymb}
\def\bkR{{\rm I\kern-.17em R}}
\def\bkC{{\rm \kern.24em \vrule width.05em height1.4ex depth-.05ex \kern-.26em C}}

\def\be{\beta}

\def\frac#1#2{{\textstyle{{#1}\over {#2}}}}

\def\lsim{\mathrel{\rlap{\lower4pt\hbox{\hskip1pt$\sim$}}
    \raise1pt\hbox{$<$}}}
\def\gsim{\mathrel{\rlap{\lower4pt\hbox{\hskip1pt$\sim$}}
    \raise1pt\hbox{$>$}}}
\def\sqr#1#2{{\vcenter{\vbox{\hrule height.#2pt
         \hbox{\vrule width.#2pt height#1pt \kern#1pt
         \vrule width.#2pt}
         \hrule height.#2pt}}}}

\def\laq{\raise 0.4 ex \hbox{$<$}\kern -0.8 em\lower 0.62 ex\hbox{$\sim$}}
\def\gaq{\raise 0.4 ex \hbox{$>$}\kern -0.7 em\lower 0.62 ex\hbox{$\sim$}}

\def\be{\begin{equation}}
\def\ee{\end{equation}}
\def\beqa{\begin{eqnarray}}
\def\eeqa{\end{eqnarray}}

\def\dalemb#1#2{{\vbox{\hrule height.#2pt
        \hbox{\vrule width.#2pt height#1pt \kern#1pt \vrule width.#2pt}
        \hrule height.#2pt}}}

\def\dalemb#1#2{{\vbox{\hrule height.#2pt
        \hbox{\vrule width.#2pt height#1pt \kern#1pt \vrule width.#2pt}
        \hrule height.#2pt}}}

\def\gtorder{\mathrel{\raise.3ex\hbox{$>$}\mkern-14mu
             \lower0.6ex\hbox{$\sim$}}}
\def\ltorder{\mathrel{\raise.3ex\hbox{$<$}\mkern-14mu
             \lower0.6ex\hbox{$\sim$}}}

\begin{document}


\title{White dwarfs in an ungravity-inspired model}

\author{Orfeu Bertolami\footnote{E-mail: orfeu.bertolami@fc.up.pt}}

\author{Hodjat Mariji\footnote{E-mail: astrohodjat@fc.up.pt}}

\vskip 0.3cm

\affiliation{Departamento de F\'\i sica e Astronomia, Faculdade de Ci\^encias da Universidade do Porto and Centro de F\'\i sica do Porto\\
Rua do Campo Alegre 687, 4169-007 Porto, Portugal}

\vskip 0.5cm

\begin{abstract}

\vskip 0.5cm

{
 An ungravity-inspired model is employed to examine the astrophysical parameters of white dwarf stars (WDs) using polytropic and degenerate gas approaches. Based on the observed properties such as mass, radius, and luminosity of selected WDs, namely, Sirius B and $\epsilon$ Reticulum, bounds on the characteristic length and scaling dimension of the ungravity (UG) model are estimated. The UG effect on the Chandrasekhar limit for WDs is shown. The UG model is examined in the study of ultra-massive WDs, e.g., EUVE J1746-706. The UG-inspired model implies that a new location for some WDs on the Hertzsprung-Russell diagram is found.
}

\vskip 0.5cm


\vskip 0.3cm

\textbf{PACS Number(s):} 04.20.Fy, 04.80.Cc, 04.25.Nx

\end{abstract}

\maketitle

\section{Introduction}

Gravity is a structural element of stellar dynamics. A change in the underlying gravity theory do have important implications for the astrophysical description (see, e.g., Refs. \cite{1} and references therein). One interesting example is $ungravity$ (UG) \cite{2}. In its pristine form, UG arises from the assumption of coupling between spin-2 unparticles and the stress-energy tensor \cite{3}. In this work, we shall consider the impact of an UG-inspired model on the astrophysics of white dwarfs (WDs). This work follows previous considerations on the effect of an UG-inspired model on the properties of the sun \cite{4}. An UG-inspired model has also been recently considered to address the flyby anomaly \cite{5}.

As will be shown, an UG model allows for the prediction of ultra-massive WDs (UWDs), i.e., WDs with masses above the Chandrasekhar limit ($M_{Ch} \simeq 1.45 M_{S} $ with $ M_{S}\simeq 2\times 10^{33} g$) such as WD 1143+321 with $M=1.52 M_{S}$ \cite {6}  \footnote{Notice that UWDs in binary systems can have their masses above $M_{Ch}$ by a small amount due to an accreted mass \cite {7}. It has also been pointed out that highly magnetized WDs can have masses as large as $M=2.58 M_{S}$, for extremely high magnetic fields 
$B_{Max} \ge 10^{13}$G \cite{Das2016}, but these are much higher than the observed magnetic fields in WDs which are typically in the ranges between $10^3$G to 
$10^9$G \cite{Ferrario2015}.}. On the other hand, these astrophysical objects allow for setting bounds on the parameters of the UG model.

In this work, the stellar equilibrium equation for WDs is obtained by considering the polytropic and degenerate gas approaches. Bounds on the UG parameters for two typical WDs, namely, Sirius B (SIB) and $\epsilon$ Reticulum (or HD 27442 B, abbreviated here by HDB) are found. The effect of UG on the Chandrasekhar mass limit of WDs is examined and UWDs such as EUVE J1746-706 is considered. Our results generalize the study the of Ref. \cite{Souza-Horvath}. Furthermore, we show how UG affects the location of a few WDs in the Hertzsprung-Russell (H-R) diagram. This paper is organized as follows: in section II, the UG model is concisely explained; in section III, the equations of the polytropic and degenerate gas models are presented; in section IV, the UG-modified equilibrium equations for WDs in the framework of both gas models are set up. Finally, our results are presented and discussed in section V.

\section{The UG Model}

The essential idea behind the UG model \cite {2} is that a modification of the Newtonian gravitational potential is introduced through the coupling of spin-2 unparticles $O^{U}_{\mu\nu}$ \cite{3} to the stress-energy tensor of Standard Model states, $T^{\mu\nu}$. The resulting stress-energy tensor has following form \cite {2}:
\begin{equation}\label{1}
  \mathcal{T}^{\mu\nu}=T^{\mu\nu}+\left( {\kappa_{*}\over \Lambda^{d_{U}-1}_{U}}\right) g^{\mu\nu}T^{\sigma\rho}O^{U}_{\sigma\rho},
\end{equation}
where $d_{U}$ and $\Lambda_{U}$ are the scaling dimension and the energy scale of $O^{U}$, respectively. In Eq. (\ref{1}), $\kappa_{*}=\Lambda^{-1}_{U}\left( \Lambda_{U}\over M_{U}\right)^{d_{UV}}$ where $M_{U}$ is the large mass scale and $d_{UV}$ is the dimension of the hidden sector operators of the ultraviolet theory which posses an infrared fixed point \cite{2}. In order to compute the effects of the unparticles to the lowest order correction to the Newtonian gravitational potential, the metric $g^{\mu\nu}$ is replaced by the Minkowski metric $\eta^{\mu\nu}$ in Eq. (\ref{1}). The resulting Newtonian gravitational potential in the UG model framework then reads \cite {2}
\begin{equation}\label{2}
 \phi_{*}(r) =-{G_{*}M\over r} \left[1+ \left( {R_{*} \over r}\right) ^{\alpha-1}\right]
\end{equation}
where $G_{*}$ is the gravitational constant of UG, $R_{*}$ is the length scale which characterizes the UG interactions, and $\alpha$ is associated with $d_{U}$ through $\alpha=2d_{U}-1$. It is obvious, from Eq. (\ref{2}), that we can recover the ordinary Newtonian gravitational potential by choosing 
\begin{equation}\label{3}
 G_{*}={G\over 1+\left( {R_{*}\over R_{0}}\right)^{\alpha-1}},
\end{equation}
where $ R_{0} $ is the distance in which the UG potential, $\phi_{*}$, matches the Newtonian one. As a good approximation, by considering the value of $\alpha$ near unity, we can write $G_{*}\simeq G/2$. Without loss of generality we set this approximation which allows for obtaining the bounds on the relevant parameters of the UG model as well as the effect of UG on the properties of WDs. Of course, the considered model is inspired on the original UG model, whose effects are expected to take place only at extremely short distances. Recent experiments set very stringent bounds on putative new interactions with ranges at submillimeter scale \cite{Tau2016} (see also Ref. \cite{OBNSantos2009} for a comparison of the results of searches of new short range interactions and the bounds for ungravity arising from nucleosynthesis considerations); however, these do not conflict with our study as we will consider deviations of the Newtonian force at range in the interval $10^{-8} R_S \lesssim R_{*} \lesssim 10^{2}R_S$, where $R_S$ is the radius of the sun.       
  
\section{The Gas Models}

A WD is considered as a mixture gas of ions and electrons which deviates from an ideal gas. The equilibrium of this compact object is ensured by the pressure, $P$, of degenerate electrons rather than by high internal temperatures as in ordinary stars. The internal temperature of WDs is rather low by stellar standards (as high as $10^{7} K$) \cite{8}. Their low luminosity ($L$), generally several orders of magnitude smaller than the one of the sun ($L_{S}= 3.846\times10^{33} erg/s $), corresponds to typical surface temperatures of order of a few times $10^{4} K$ \cite{6}. Thus, WDs are too cold to ignite nuclear reactions. Their composition at birth is mostly $^{4}He$, $^{12}C$, and $^{16}O$ (with $\mu_{e}={A\over Z}=2$). Some heavier elements can be produced during a $\textit{pycnonuclear}$ reaction process under which WDs evolve, on a very long timescale, through zero temperature nuclear reactions in which lattice vibrations yield a small, but finite probability of Coulomb barrier tunneling \cite{6, 9}. In this work, we assume that a WD is at zero-temperature and behaves as a neutral gas of non-interacting electrons and bounded nucleons in nuclei of which the composition parameter is $\mu_{e}=2$. The electrons, whether they are relativistic or not, contribute virtually to the entire pressure of the WD, while the bounded nucleons contribute virtually to all the WD energy density, given by $\mu_{e}m_{H}c^{2}n_{e}$, where $m_{H}$ is the atomic mass of the hydrogen ion, and $n_{e}$ as the density of electrons. We assume either the polytropic or the degenerate gas models to establish the Newtonian hydrostatic equilibrium (NHE) equation, for the WDs. In Sec. IV, the validity of the NHE equation for WDs will be investigated. For a static Newtonian star, the NHE equation is given by \cite{8}: 
\begin{equation}\label{4}
 {dP(r)\over dr} = -{ GM(r)\rho(r)\over r^{2} },
\end{equation}
where a further derivative with respect to $r$ leads to the usual form of NHE equation:
\begin{equation}\label{5}
 {1\over r^{2}}{d \over dr}\left({r^{2} \over \rho}{dP(r) \over dr}\right) =-4\pi G\rho(r).
\end{equation}

Next, we consider the two gas models.

\subsection{Polytropic gas model}
  
  According to the polytropic gas model, the pressure depends on the density, $\rho$, as follows \cite{8}:
\begin{equation}\label{6}
 P=K \rho^{(n+1)/n},
\end{equation}
where $n$ is the polytropic index and $K$ is a constant factor. With this equation of state (EoS), we can obtain the well-known form of the Lane-Emden (LE) equation. In order to do this, we introduce two dimensionless variables, $\theta$, and, $\xi$, to express the density and radial distance with respect to the center of star values, respectively:
\begin{equation}\label{7}
 \rho=\rho_{c} \theta^{n},
\end{equation}
\begin{equation}\label{8}
r =\beta_{p} \xi,
\end{equation}
where $\rho_{c}$ is the density at the center of a star and $\beta_{p}=\left [ \frac{\left( n+1\right) K}{4\pi G}\rho_{c}^{(1-n)/n} \right]^{1/2} $. The pressure of a polytropic gas reads
\begin{equation}\label{9}
 P=P_{c} \theta^{n+1},
\end{equation}
where $P_{c}=K \rho_{c}^{(n+1)/n}$. Substituting Eqs. (\ref{7})-(\ref{9}) into Eq. (\ref{5}) yields the well known LE equations:
\begin{equation}\label{10}
 {1 \over \xi^{2}} {d \over d\xi}\left(\xi^{2} {d\theta\over d\xi}\right)=-\theta^{n}.
\end{equation}
The above differential equation should be solved submitted to the following boundary conditions: $\theta(\xi=0)=1$ and $\theta^{\prime}(\xi=0)=0$. The density and pressure of the star can be obtained through solution of the LE equation for each value of $\xi$. The first zero of the LE equation solutions (the value of $\theta(\xi)=0$ for the first zero, indicated as $\xi_{10}$) allows for determining the relevant quantities of a star, such as its radius and mass. The radius of a star is obtained as
\begin{equation}\label{11}
 R=\beta_{p}\xi_{10}.
\end{equation}
Using relation $dM(r)=4\pi\rho(r)r^{2}dr$, together with Eqs. (\ref{7}) and (\ref{8}), as well as the LE equation, leads to
\begin{equation}\label{12}
 M(\xi_{10})= 4\pi \rho_{c} \beta_{p}^{3}(-\xi^{2} {d\theta\over d\xi})\mid_{\xi=\xi_{10}}.
\end{equation}
Finally, eliminating $\rho_{c}$ in Eq (\ref{12}), we obtains a relation between the mass and the radius of the star \cite{8}
\begin{equation}\label{13}
4\pi M^{n-1}R^{3-n}=\left[ {(n+1)K\over G}\right]^{n}\left[ \left( -{d\theta_{n}\over d\xi}\right)_{\xi_{10}}\right]^{n-1} \left( \xi_{10} \right)^{n+1}.  
\end{equation}

\subsection{Degenerate gas model}
   
We assume now that WDs are completely described as a electron-degenerate gas with densities in the range of $10^{5}-10^{8} g/cm^{3}$ \cite{8}. On the other hand, WDs satisfy the degeneracy condition in which the temperature should be much smaller than the Fermi energy $E_{F}=\sqrt{p_{F}^{2}c^{2}+E_{0e}^{2}}$ where $p_{F}$ is the Fermi momentum and $E_{0e} \simeq 8\times 10^{-6} erg$ $(T\simeq 6\times10^{9} K)$, the rest energy of electrons. With this assumption, the electron density distribution function can be given approximately by the Heaviside function and the ensued the electron density as follows: 
\begin{equation}\label{14}
n_{e}={1\over \pi^{2}\hbar^{3}}\int _{0}^{p_{F}}p^{2}dp={E_{0e}^{3}\over 3\pi^{2}(\hbar c)^{3}}x^{3},
\end{equation}
where $\hbar$ is the Planck constant and $ x=p_{F}/m_{e}c$.
The pressure of electron gas is given by \cite{8}
\begin{equation}\label{15}
P={1 \over 3\pi^{2}\hbar^{3}}\int_{0}^{p_{F}}{p^{2} \over \sqrt{m^{2}+{p^{2}\over c^{2}}}}p^{2}dp=Af(x),
\end{equation}
where $A=E_{0e}^{4}/24\pi^{2}(\hbar c)^{3}\simeq 6.002\times10^{22} erg/cm^{3}$ and
\begin{equation}\label{16}
f(x)=x(2x^{2}-3)(x^{2}+1)^{1/2}+3sinh^{-1}(x).
\end{equation}
As the gas is neutral, through Eq. (\ref{14}) we can write the density of the WD in the degenerate gas model as \cite{8}: 
\begin{equation}\label{17}
\rho=Bx^{3},
\end{equation}
where $B=E_{0e}^{3}\mu_{e}m_{H}/3\pi^{2}(\hbar c)^{3}\simeq9.74\times10^{5}\mu_{e}$ $g/cm^{3}$. This equation, together with Eq. (\ref{15}) are known as the EoS of WDs in the framework of completely electron-degenerate gas model.  
 In order to obtain the NHE equation in the degenerate gas model, we substitute Eqs. (\ref{15}), (\ref{16}), and (\ref{17}) into Eq. (\ref{5}) to obtain
\begin{equation}\label{18}
 {1\over r^{2}}{d\over dr}\left(r^{2}{dX\over dr}\right) =-{\pi GB^{2}\over 2A}x^{3}
\end{equation}  
where $X=\sqrt{x^{2}+1}$. By defining the new variable $\Phi$ as
\begin{equation}\label{19}
 X=X_{c}\Phi,
\end{equation}
where $X_{c}$ is the value of $X$ at the center of star, and $\xi=r/\beta_{d}$ with $\beta_{d}=\sqrt{2A\over \pi GB^{2}X_{c}^{2}}$. The NHE equation then reads
\begin{equation}\label{20}
{1\over \xi^{2}}{d\over d\xi}\left(\xi^{2}{d\Phi\over d\xi}\right)=-(\Phi^{2}-X_{c}^{2})^{\frac{3}{2}}.
\end{equation}
This LE equation for a degenerate gas, can be solved once $\rho_{c}$ is known and boundary conditions specified: $\Phi(\xi=0)=1$ and $\Phi^{\prime}(\xi=0)=0$. In contrast to Eq. (\ref{10}), $ \xi_{10}$ is the first zero of Eq. (\ref{20}) so that $X(\xi_{10})=1 $. The radius of a WD is obtained as
\begin{equation}\label{21}
 R=\beta_{d}\xi_{10}=7.77\times10^{8}{1\over \mu _{e}X_{c}}\xi_{10}.
\end{equation}
Similarly, the mass of WD can be obtained by
\begin{equation}\label{22}
 M(\xi_{10})= 4\pi BX_{c}^{3}\beta_{d}^{3}(-\xi^{2} {d\Phi\over d\xi})\mid_{\xi_{10}}.
\end{equation}

\section{The LE Equation for the UG Model}

In order to study the effect of UG on WDs, we must suitably adjust the LE equation. In this work, we use a method similar to the one of Refs. \cite{1, 4} to obtain the modified LE equation for both polytropic and degenerate gas models. We first argue that the NHE equation is a valid approximation of the most general Tolman-Oppenheimer-Volkoff (TOV) equation for a WD \cite{10}:
\begin{equation}\label{23}
 4\pi r^{2}dP(r)=-{GM(r)dM(r)\over r^{2}}\left[1+{P(r)\over \rho(r)c^{2}}\right]\left[1+{4\pi r^{3}P(r)\over M(r)c^{2}}\right]\left[1-{2GM(r)\over c^{2}r}\right]^{-1},
\end{equation}
where $dM(r)=4\pi r^{2}\rho(r) dr$. For WDs, $P(r)\ll\rho(r)c^{2}$ in the non-relativistic (low density) and the ultra-relativistic (high density) limits. In order to show this, we focus on the EoS of WDs in the degenerate gas model (subsection B) at the two limits. In the non-relativistic limit, $p_{F}c \ll E_{0e}$ or equivalently $x\ll 1$, hence $f(x)\sim{8\over 5}x^{5}$. Thus, from Eqs. (\ref{15}) and (\ref{17}), we can write
\begin{equation}\label{24}
{P\over \rho c^{2}}\sim {8A\over 5Bc^{2}}x^{2}\simeq 5\times10^{-5}x^{2} \ll 1.
\end{equation}
Therefore, at the non-relativistic regime or at the low density regions ($x\ll 1$) we have $P\ll \rho c^{2}$. At the ultra-relativistic limit, $p_{F}c\gg E_{0e}$ or equivalently $x\gg 1$, the expansion of $f(x)$ can be approximated by $\sim2x^{4}$ and then
\begin{equation}\label{25}
{P\over \rho c^{2}}\sim {2A\over Bc^{2}}x\simeq 7\times10^{-6}\left( {p_{F}c\over E_{0e}}\right) \ll 1.
\end{equation}
Indeed, the density of WD is $10^{9} g/cm^{3}$, considering $\rho=\mu_{e}m_{H}n_{e}$, the Fermi momentum of electron, $k_{Fe}=(3\pi^{2}n_{e})^{1/3}$, is about $0.045 fm^{-1}$ and $p_{F}c/{E_{0e}} \sim 17$, hence, from Eq. (\ref{25}), $P\ll \rho c^{2}$.
Thus, we can neglect the second term in the first bracket in Eq. (\ref{23}). Regarding the second bracket in Eq. (\ref{23}), rearranging $ M(r)\sim 4\pi r^{3}\overline{\rho}/3$, where $\overline{\rho}$ is the average density up to radius $r$, then $4\pi r^{3}P\sim3(P/\overline{\rho} c^{2})M(r)c^{2}\ll M(r)c^{2}$, and hence we can ignore the second term of the second bracket of Eq. (\ref{23}). Finally, for the third bracket, as no region of the star lies within its Schwarzschild radius, $\frac{2GM(r)}{rc^{2}}\ll 1$.

We now consider the UG hydrostatic equilibrium (UGHE) equation. We incorporate the UG-modified Newtonian gravitational potential, Eq. (\ref{2}), in the NHE equation, Eq. (\ref{4}), as follows
\begin{equation}\label{26}
 {dP(r)\over dr}=-{G_{*}M(r)\rho(r)\over r^{2}}\left[1+\left( {R_{*}\over r}\right) ^{\alpha-1}\right].
\end{equation}
By employing $dM(r)=4\pi \rho(r)r^{2}dr$, after a straightforward calculation, the UGHE equation becomes
\begin{equation}\label{27}
 {1\over r^{2}}{d\over dr}\left({r^{2}\over \rho}{dP(r)\over dr}\right) =-4\pi G_{*}\rho(r)\left[ 1+\alpha\left( {R_{*}\over r}\right)^{\alpha-1}\right] + {G_{*}M(r)\over R_{*}^{3}}\left[ \alpha(\alpha-1)\left( {R_{*}\over r}\right)^{\alpha+2}\right] .
\end{equation}
It is clear that setting $\alpha=1$ and $G_{*}=G/2$ in the UGHE equation leads to the NHE equation, Eq. (\ref{4}).

In order to obtain LE-modified equation, we include the EoS of both gas models, i.e., Eqs. (\ref{7}) and (\ref{8}) for the polytropic gas model and Eqs. (\ref{15}) and (\ref{17}) for the degenerate gas model, in the UGHE equation. From Eq. (\ref{12}) we obtain, after some manipulation, the modified LE equation for the polytropic gas model:
\begin{equation}\label{28}
 {1\over \xi^{2}}{d\over d\xi}\left(\xi^{2}{d\theta\over d\xi}\right)=-{G_{*}\over G}\left\lbrace \left[ 1+\alpha\left( {\xi_{*}\over \xi}\right)^{\alpha-1}\right]\theta^{n}+\left[\alpha(\alpha-1)\left( {\xi_{*}\over \xi}\right)^{\alpha-1} \left( {1\over \xi} {d\theta \over d\xi}\right) \right]\right\rbrace.
\end{equation}
From Eq. (\ref{22}) we get, after some manipulation, the modified LE equation for the degenerate gas model:
\begin{equation}\label{29}
{1\over \xi^{2}}{d\over d\xi}\left(\xi^{2}{d\Phi\over d\xi}\right)=-{G_{*}\over G}\left\lbrace \left[ 1+\alpha\left( {\xi_{*}\over \xi}\right)^{\alpha-1}\right](\Phi^{2}-X_{c}^{2})^{\frac{3}{2}}+ \left[\alpha(\alpha-1)\left( {\xi_{*}\over \xi}\right)^{\alpha-1} \left( {1\over \xi} {d\Phi \over d\xi}\right) \right]\right\rbrace.
\end{equation}
In Eqs. (\ref{28}) and (\ref{29}), $\xi_{*}=R_{*}/\beta_{p(d)}$ for the polytropic (degenerate) gas. Choosing $\alpha=1$ and $G_{*}=G/2$, we recover the usual LE equations, Eqs. (\ref{10}) and (\ref{20}). The mass and radius of WDs are calculated by Eqs. (\ref{11}) and (\ref{12}) for the polytropic gas model or by Eqs. (\ref{21}) and (\ref{22}) for the degenerate gas model at $\xi^{*}_{10}$, the first zeros of the modified LE equations.

\section{Results and Discussion}

We consider the UG model for two arbitrary WDs, i.e., SIB and HDB, in the framework of polytropic and degenerate gas models. Table I indicates the values of the mass ($M_{0}$), radius ($R_{0}$), and luminosity ($L_{0}$) in terms of the corresponding parameters of the sun ($M_{S}$, $R_{S}$, and $L_{S}$), along with data of their effective temperatures. The data of $M_{0}$, $R_{0}$, and $T_{eff}$ arise from the gravitational redshift method, as quoted by Refs. \cite{11, 12}. The luminosity, $L$, is given by
\begin{equation}\label{30}
L=4\pi R^{2}\sigma T_{eff}^{4},
\end{equation}     
where $\sigma$ is Stefan-Boltzmann constant. Regarding the values of the effective temperature and radius of SIB and HDB, from the corresponding uncertainties, we obtain $L_{0}$ and $\triangle L_{0}$.

\vspace{.7 cm}
\begin{center}
{\footnotesize Table I Relevant values for the selected WDs, i.e., SIB and HDB \cite{11, 12}.}
\\
\vspace{.5 cm}
\begin{tabular}{|c|c|c|c|c|}
\hline
WD & $(M_{0}\pm\triangle M_{0})/M_{S}$ & $(R_{0}\pm\triangle R_{0})/R_{S}$ & $ T_{eff}\pm\triangle T_{eff} (K)$ & $(L_{0}\pm\triangle L_{0})/L_{S}$\\
\hline
 SIB & 1.02$\pm$0.02 & 0.0081$\pm$0.0002 & 25193$\pm$37 & 0.0237$\pm$0.0013 \\
 HDB & 0.616$\pm$0.022 & 0.0129$\pm$0.0003 & 15310$\pm$350 & 0.0082$\pm$0.0011\\
\hline
\end{tabular}
\end{center}
\vspace{.7 cm}

Our method consists in using the uncertainties of the relevant quantities to obtain bounds on the characteristic length, $R_{*}$, and scaling dimension, $\alpha$, of the UG inspired-model.
In order to compute the astrophysical bounds on $\alpha$ and $R_{*}$ and to get the new mass limit for WDs, we outlines the adopted strategy. At first, we solve the LE equations, Eqs. (\ref{10}) and (\ref{20}), to obtain for the selected WDs mass, radius, and luminosity, denoted by $M_{10}$, $R_{10}$, and $L_{10}$, respectively. Then, by varying $\alpha$ and $R_{*}$ within the LE-modified equations, Eqs. (\ref{28}) and (\ref{29}), we calculate the same observable parameters and accept those values that are compatible with the uncertainties (Table I). Next, the effect of UG on the Chandrasekhar limit mass is examined. Finally, we depict the effect of UG on the position of a few WDs in H-R diagram.

We set the polytropic index of $n=2.03 (1.73)$ and the core density $\rho_{c}=3.20\times 10^{7} (3.22\times10^{6}) g/cm^{3}$ for SIB (HDB). Table II shows the calculated mass, radius, and luminosity. 

\vspace{.7 cm}
\begin{center}
{\footnotesize Table II The computed values of the properties of the selected WDs (SIB and HDB).}
\\
\vspace{.5 cm}
\begin{tabular}{|c|c|c|c|c|}
\hline
Model & WD & $M_{10}/M_{S}$ & $ R_{10}/R_{S}$ & $ L_{10}/L_{S}$ \\
\hline
Degenerate & SIB & 1.0988 & 0.0080 & 0.0231  \\
 & HDB & 0.6012 & 0.0127 & 0.0079  \\
\hline
Polytropic & SIB & 1.0201 & 0.0081& 0.0237  \\
 & HDB & 0.6162 & 0.0129 & 0.0082  \\
\hline
\end{tabular}
\end{center}
\vspace{.7 cm}

\begin{figure}
\centering
\hfill
\subfigure [HDB]{\includegraphics[width=.4\linewidth]{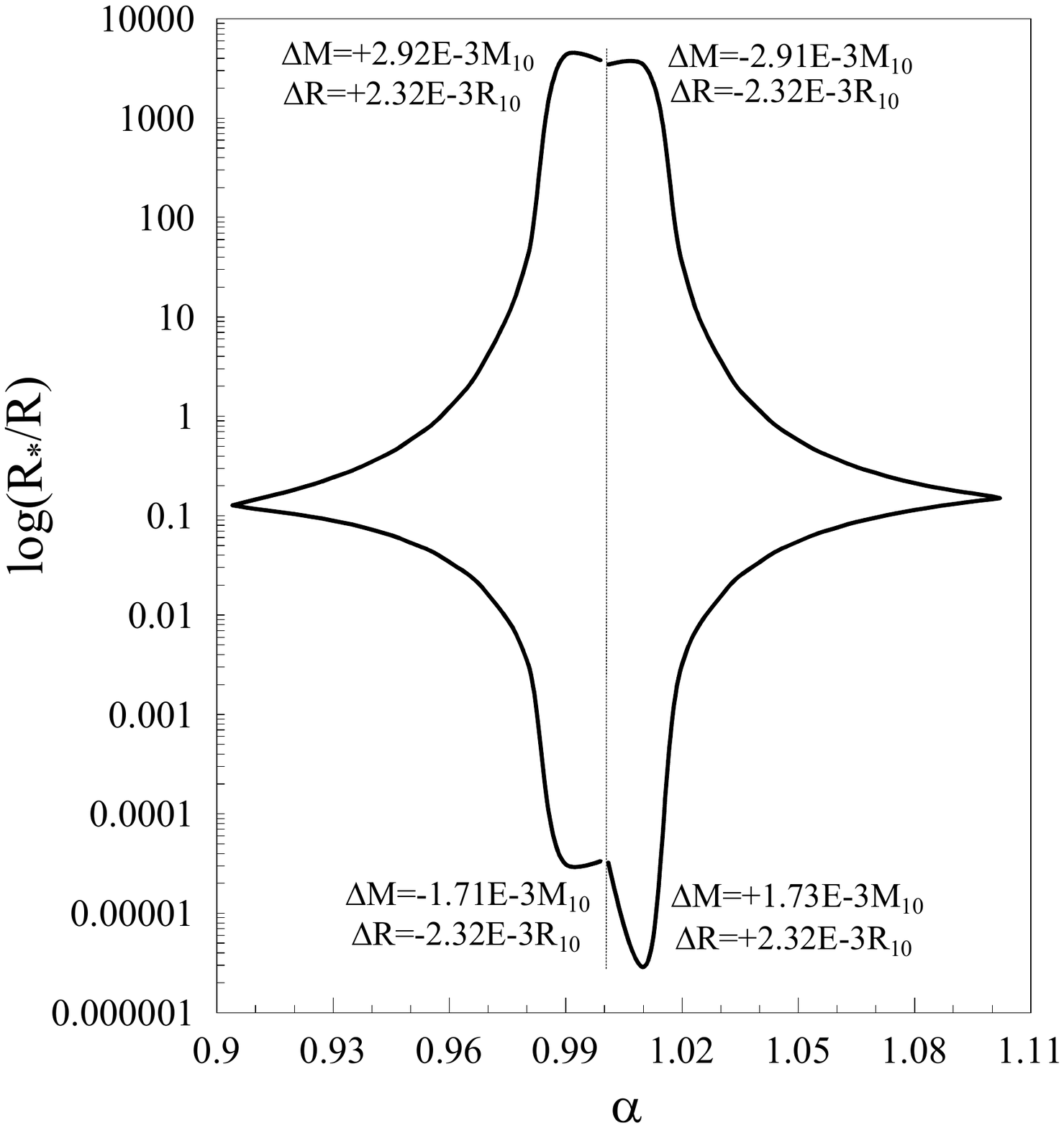}}
\hfill
\subfigure [SIB]{\includegraphics[width=.4\linewidth]{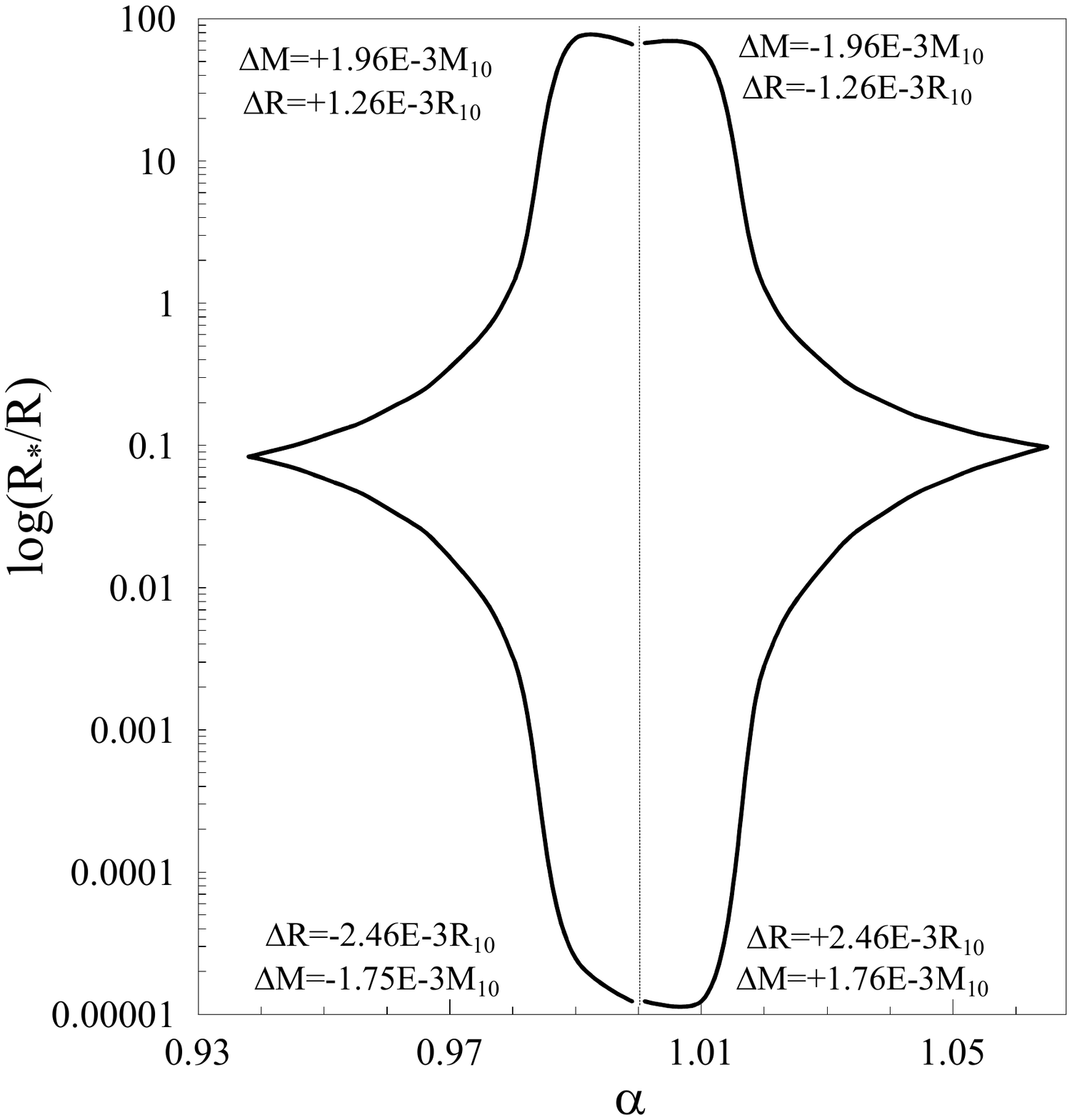}}
\hfill
\hfill
\caption{The allowed region for the UG parameters for (a) HDB and (b) SIB with the polytropic gas model. The characteristic length has been normalized by $R$, the radius of the relevant WD.}\label{Fig:Pol}
\end{figure}

\begin{figure}
\centering
\hfill
 \subfigure [HDB] {\includegraphics[width=.4\linewidth]{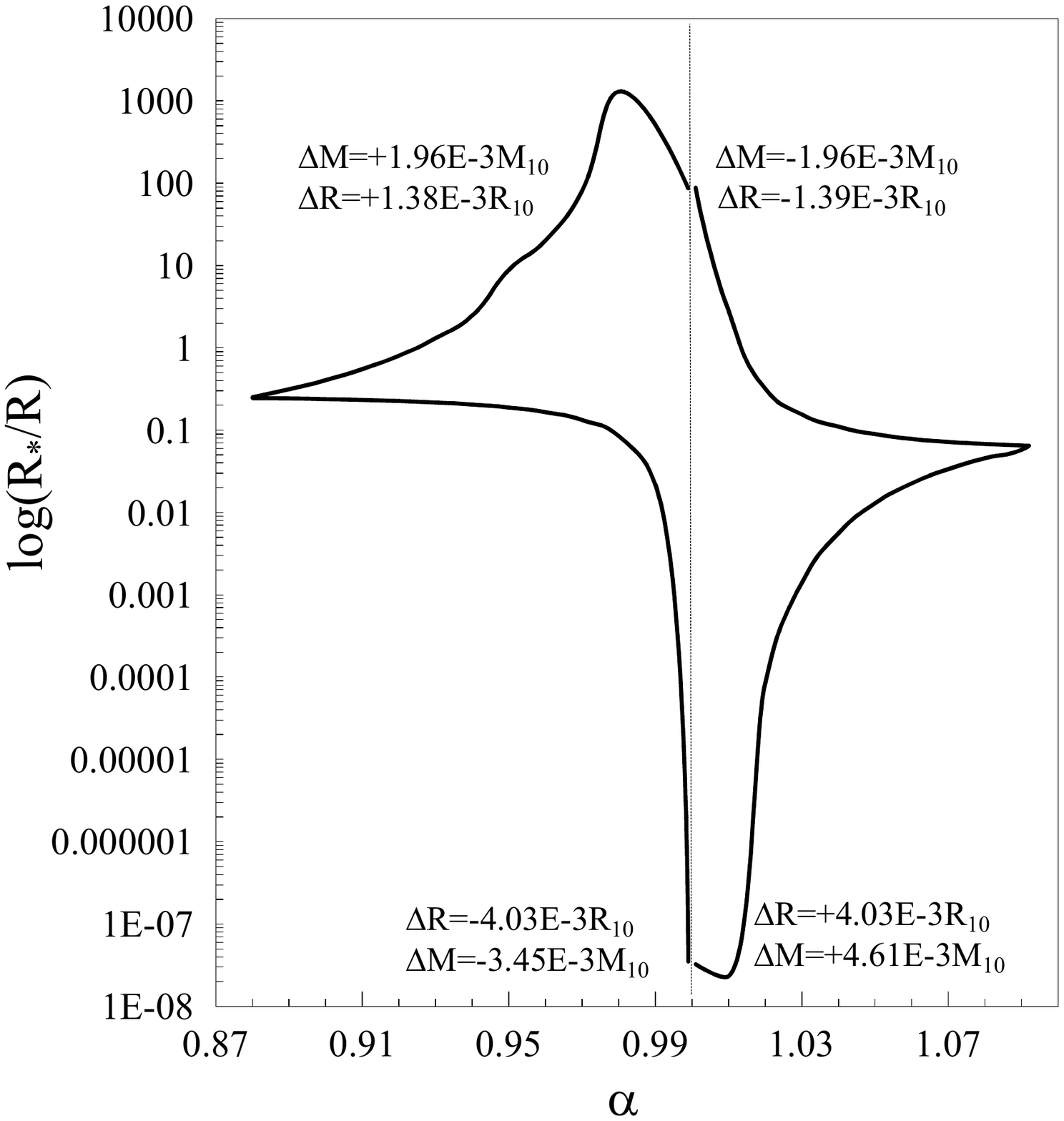}}
\hfill
 \subfigure [SIB] {\includegraphics[width=.4\linewidth]{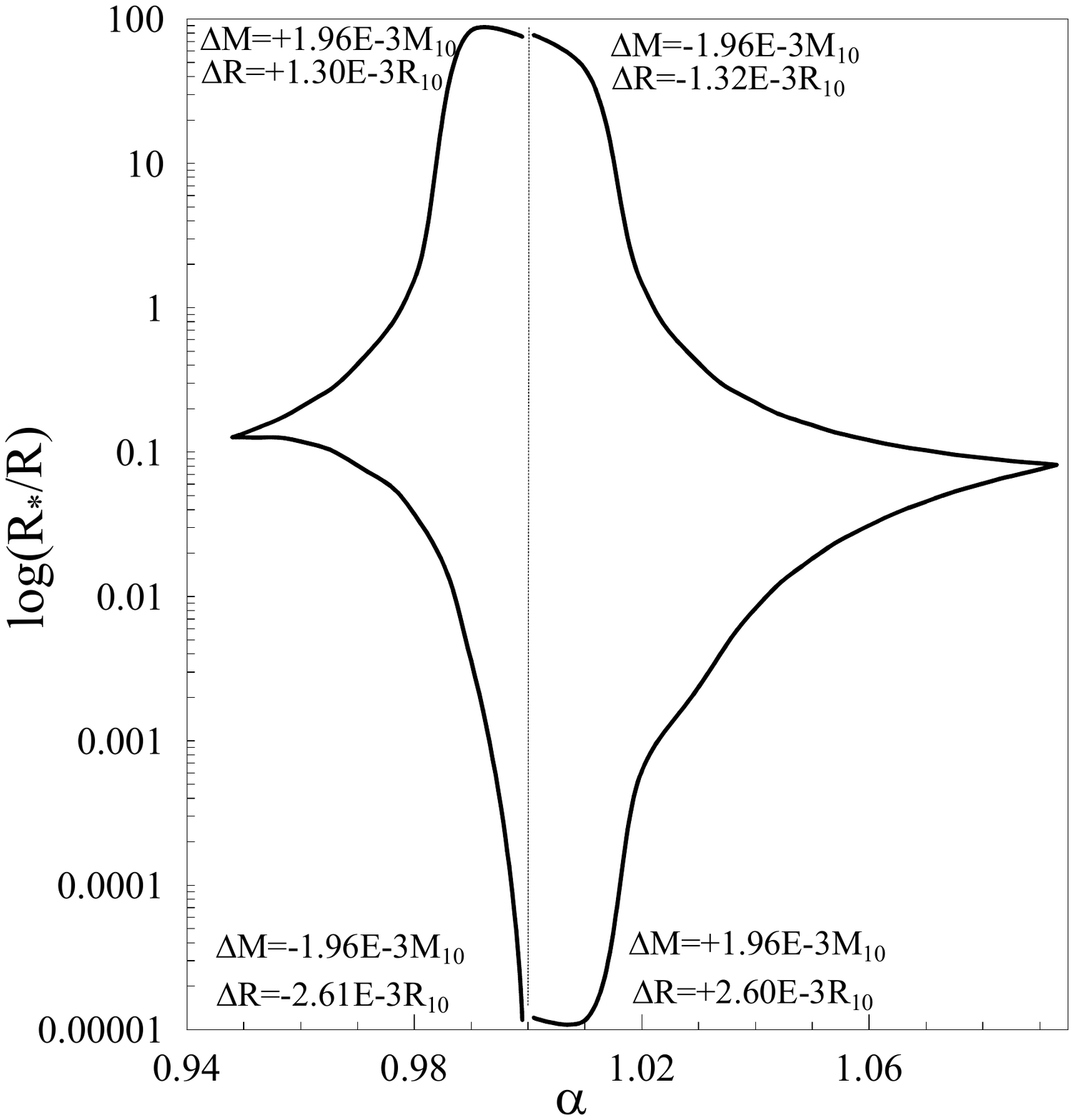}}
\hfill
\hfill
\caption{The same as Fig. 1 for the degenerate gas model.}\label{Fig:Deg}
\end{figure}

For the same input parameters, that is, $\rho_{c}$ and $n$, we solve Eqs. (\ref{28}) and (\ref{29}) for the different values of $\alpha$ and $R_{*}$. We select those solutions for which $M$, $R$, and $L$, calculated at $\xi^{*}_{10}$, remain within the observational range as illustrated by Table I, i.e., $\left[ M_{0}-\bigtriangleup M_{0},M_{0}+\bigtriangleup M_{0}\right] $, etc. In order to find the allowed region for $R_{*}$ and $\alpha$, we computed the upper and lower bounds on $R_{*}$ denoted by $R_{*}^{+}$ and $R_{*}^{-}$, respectively. Figs.  ~\ref{Fig:Pol} and ~\ref{Fig:Deg} depict the allowed regions of $R_{*}$ and $\alpha$ based on the degenerate and polytropic gas models, for HDB (SIB) (panels (a(b))). In order to obtain $R_{*}^{+}(R_{*}^{-})$, we use the upper (lower) values of $M$ so that the values of $R$ and $L$ remain within the observational range (cf. Table I). It should be mentioned that in each portion of the allowed regions we set a fixed value for the uncertainty in $M$, $R$, and $L$. From Eqs. (\ref{11}), (\ref{12}), (\ref{21}), (\ref{22}), and (\ref{30}) we obtain
\begin{equation}\label{31}
\bigtriangleup R=\left[ \left( {\xi^{*}_{10}\over \xi_{10}}\right)-1\right] R_{10},
\end{equation}  
for the uncertainty in $R$,
\begin{equation}\label{32}
\bigtriangleup L=\left[ \left( {\xi^{*}_{10}\over \xi_{10}}\right)^{2}-1\right] L_{10},
\end{equation}  
for the uncertainty in $L$, and
\begin{equation}\label{33}
\bigtriangleup M=\left[ \left( {\xi^{*}_{10}\over \xi_{10}}\right) ^{2}\left({\eta^{\prime}_{11}\over \eta^{\prime}_{10}}\right)-1\right] M_{10},
\end{equation}
for the uncertainty in $M$. In Eq. (\ref{33}), $\eta^{\prime}$ indicates $\theta^{\prime}(\Phi^{\prime})$, the derivative of the LE solution for the polytropic (degenerate) gas model. The $\bigtriangleup M$ and $\bigtriangleup R$ values shown in Figs.~\ref{Fig:Pol} and ~\ref{Fig:Deg} are around $\alpha=1$. Table III shows the astrophysical bounds on $\alpha$ and $R_{*}$ with respect to data of SIB and HDB.

\vspace{.5 cm}
\begin{center}
{\footnotesize Table III The astrophysical bounds on $\alpha$ and $R_{*}$ with respect to the sample WDs' data.}
\\
\vspace{.5 cm}
\begin{tabular}{|c|c|c|c|c|c|c|}
\hline
Model & WD & $\alpha$ & $ R_{*}(m)$ & $M/M_{S}$ & $R/R_{S}$ & $L/L_{S}$\\
\hline
Degenerate & SIB & 0.948 & 713.707 & 1.040 & 0.0079 & 0.0226 \\
                && 1.093 & 460.951 & 1.000 & 0.0083 & 0.0248 \\
           & HDB & 0.880 &2261.582 & 0.638 & 0.0126 & 0.0078 \\
                && 1.092 & 581.410 & 0.594 & 0.0132 & 0.0858 \\
\hline
Polytropic & SIB & 0.942 & 445.632 & 1.038 & 0.0079 & 0.0226 \\
                && 1.065 & 550.077 & 1.000 & 0.0083 & 0.0248 \\
           & HDB & 0.904 &1141.932 & 0.638 & 0.0126 & 0.0078 \\
                && 1.102 &1345.948 & 0.594 & 0.0132 & 0.0858 \\
\hline
\end{tabular}
\end{center}
\vspace{.7 cm}

From values in Table III, we see that the allowed values of the UG parameters decrease when the central density, or equivalently the ratio $M/R$, increases. For example, in the framework of the polytropic model, when the core density increases an order of magnitude, $\alpha$ gets closer to unity by about $4$ percent. A stronger behavior is found for $R_{*}$. For instance, based on the limit values of $\alpha$ for the polytropic model, $R_{*}$ gets reduced by sixty percent with a tenfold increase in the density. As a result, by increasing the $M/R$, the allowed region for the UG parameters becomes smaller. 

\begin{figure}
\begin{center}

\includegraphics[scale=0.6]{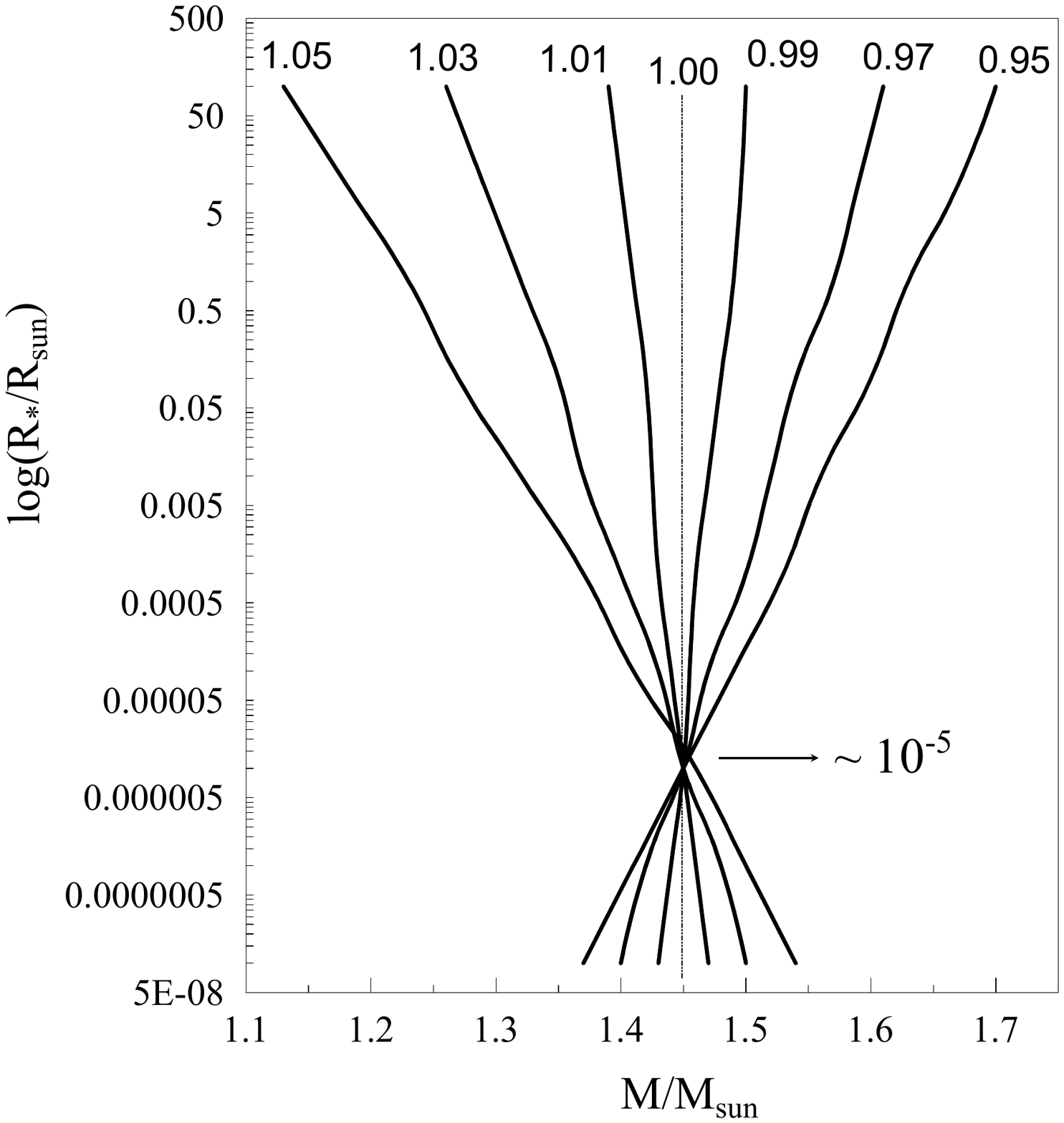}

\caption{The characteristic length of UG vs. the mass limit of WDs for different $R_{*}$ and $\alpha$ values.}\label{Fig:UG-MR}
\end{center}
\end{figure}

We now estimate the effect of UG on the Chandrasekhar mass limit, $M_{Ch}$. At the ultra-relativistic limit, $x\gg 1$, from Eq. (\ref{16}), $f(x)\sim 2x^{4}$. Hence, using Eqs. (\ref{15}) and (\ref{17}), we can write
\begin{equation}\label{34}
P=\left( {2A\over B^{4/3}}\right) \rho^{4/3}.
\end{equation}
This EoS corresponds to a polytropic gas with $n=3$. With the values of $A$, $B$, and $\mu_{e}$, using Eq. (\ref{22}), the Chandrasekhar mass limit reads
\begin{equation}\label{35}
M_{Ch}=0.721\left(-\xi^{2} \theta^{\prime}\right)\mid_{\xi_{10}}M_{S},
\end{equation} 
where $\theta^{\prime}$ indicates the derivative of the LE solution for the ultra-relativistic polytropic gas model. Hence, from the value of $\xi_{10}=6.89679$ and $\theta^{\prime}_{10}=-0.04243$, we obtain the well known result, $M_{Ch}=1.45 M_{S}$. When we switch on UG, the value of the first zero of the modified LE equation and of the corresponding derivative are changed and thus we can obtain new mass limits for WDs as a function of $\alpha$ and $R_{*}$. Fig.~\ref{Fig:UG-MR} illustrates how the mass limit of WDs varies with $R_{*}$ for different $\alpha$'s. As depicted in Fig.~\ref{Fig:UG-MR}, it is possible to have WDs with masses greater than $M_{Ch}$ for different values of $\alpha$ and $R_{*}$. As mentioned in Sec. I, the mass of WD 1143+321 is higher than $M_{Ch}$ ($M=1.52 M_{S}$ \cite {6}). Thus, the existence of this WD can be accommodated within the UG model. 
As shown in Fig.~\ref{Fig:UG-MR}, the curves get closer to ordinary gravity case when $\alpha \rightarrow 1^{\pm}$. Actually, it can be seen that the curves rotate clockwise (counter clockwise) around a point with $M=M_{Ch}$ and $R_{*}\sim 10^{-5} R_{S} \simeq 7  km$ for $\alpha \rightarrow 1^{+(-)}$ (but for $\alpha = 1.05$). It means for $\alpha \rightarrow 1^{\pm}$ we can recover the usual Chandrasekhar limit mass independently of the characteristic length of UG for $R_{*}\simeq 7 km$. This is achieved without any extra assumption beyond the choice $n=3$, $\rho_{c}\simeq 10^{10} g/cm^{3}$ and the ordinary boundary conditions to solve the LE equation, Eq. (\ref{28}). Notice that Fig.~\ref{Fig:UG-MR} shows for $\alpha=1$, that UG-inspired model also predicts that the mass limit for WDs is smaller than the usual value. Fig.~\ref{Fig:UG-MR} and the corresponding data might be thus observationally useful.

\begin{figure}
\begin{center}

\includegraphics[scale=0.6]{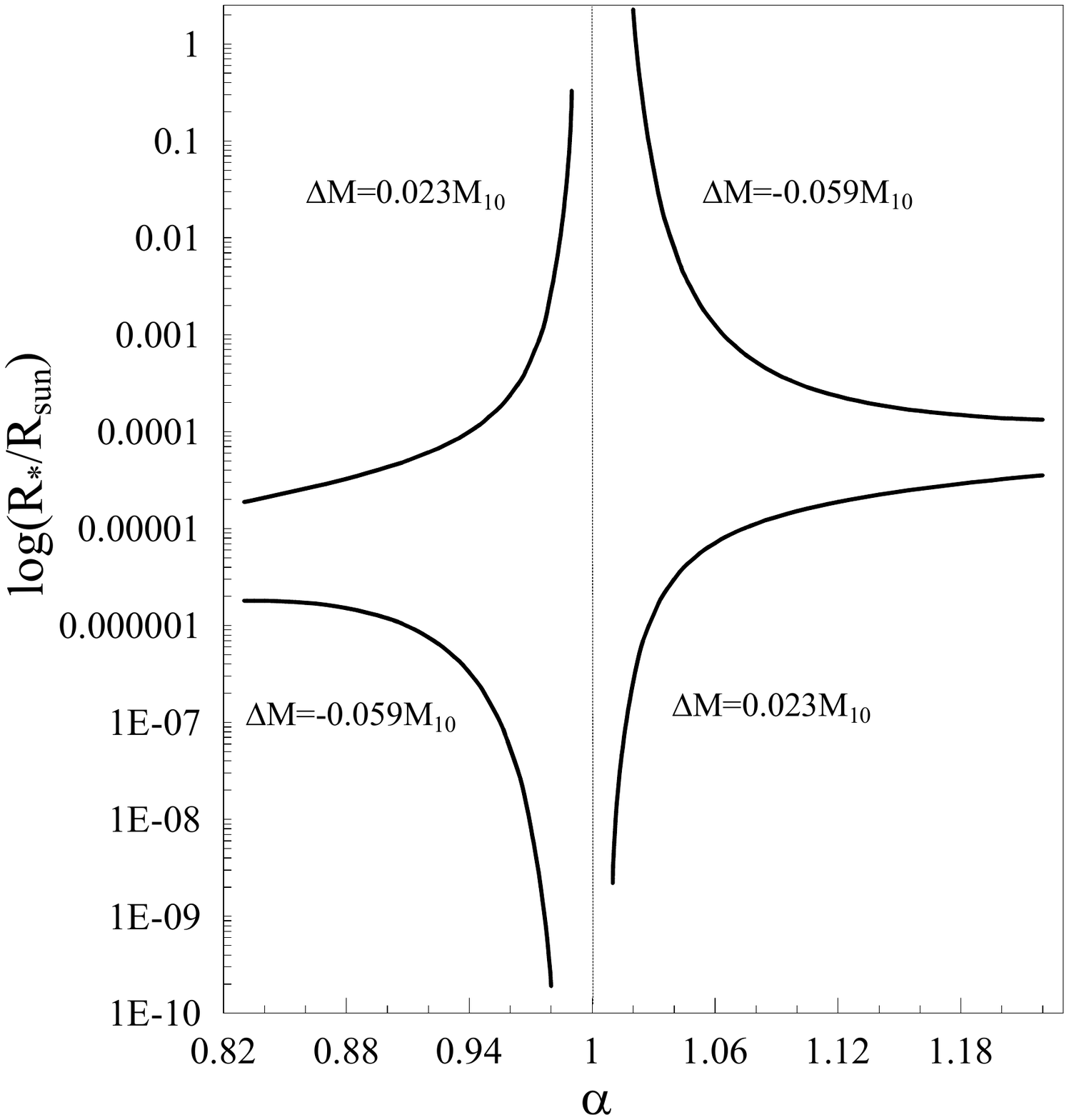}

\caption{The allowed region for UG for the EUVE J1746-706 WD, using the polytropic gas model.}\label{Fig:EUVE706}

\end{center}
\end{figure}

Although UWDs ($M > 1.1 M_{S}$) are rather rare with respect to the ordinary WDs ($M\sim0.6 M_{S}$), they can be observed through  gravitational redshift measurements, radius estimates or surface gravity measurements \cite{13}, obtained, for instance, by surveys of the Extreme Ultraviolet Explorer (EUVE) \cite{14}. We apply the UG model on an UWD, namely, EUVE J1746-706. According to the observational data, $M=1.43M_{S}$, and $\bigtriangleup M=0.06 M_{S}$ \cite{14}, and for the UG polytropic gas model ($n=3$), the $\alpha-R_{*}$ plot dependence is shown in Fig.~\ref{Fig:EUVE706}. As depicted in Fig.~\ref{Fig:EUVE706}, the tail of the curves is longer than the ordinary WDs. Although it seems that the curves in both region $\alpha < 1$ and $\alpha > 1$ do not meet each other unless for $\alpha$ very far from unity, we expect this behavior for curves of UWDs since we use NHE, Eq. (\ref{4}), to get the modified LE equation, Eq. (\ref{28}). We envisage that including general relativity corrections on UGHE might be led to a reliable bounds on the UG parameters for UWDs. It is worth mentioning that the obtained bounds for $R_{*}$ and $\alpha$ are compatible with the ones obtained from the UG LE equation solutions applied for the sun using the $6$ percent uncertainty on its core temperature \cite{4}.
 
\begin{figure}
\begin{center}

\includegraphics[scale=0.6]{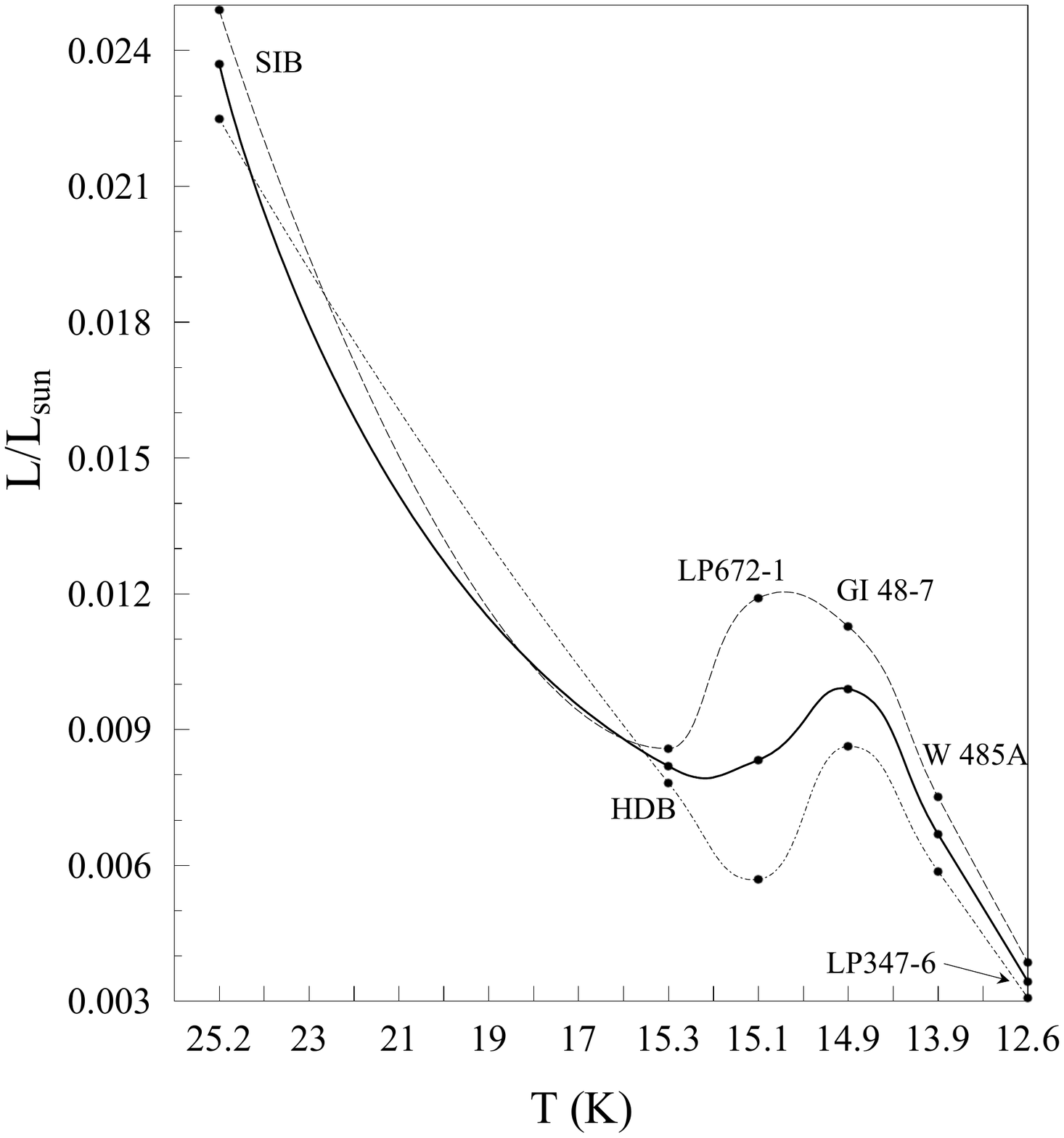}

\caption{The H-R diagram for a few WDs \cite{11, 12, 15}. The solid, dashed, and dash-dotted curves corresponded to $\alpha=1$, $\alpha>1$, and $\alpha<1$, respectively.}\label{Fig:LTPoln2}

\end{center}
\end{figure}

 At the final step, we show how UG changes the location of WDs in the H-R diagram. In order to do this, we obtain the bound values of $R_{*}$ and $\alpha$ for a few WDs with respect to their mass and radius and the corresponding uncertainties \cite{11, 12, 15} and compute their luminosity. Table IV shows the bounds on $R_{*}$ and $\alpha$ by considering the observational data. It should be pointed out that the calculations are performed in the framework of the polytropic model with $n=2$. The luminosity of the selected WDs can be computed by knowing their radius and surface temperature. Fig. ~\ref{Fig:LTPoln2} illustrates the position of WDs in the H-R diagram for $\alpha=1$ (solid curve), $\alpha>1$ (dashed curve), and  $\alpha<1$ (dash-dotted curve). It is clear that all curves for different values of $\alpha$ and $R_{*}$ are between the dashed and dash-dotted curves. Once again we can see the role played by UG on the determination of luminosity of WDs.

\vspace{10 cm}
\begin{center}
{\footnotesize Table IV The astrophysical bounds on $\alpha$ and $R_{*}$ with respect to the selected WDs' data \cite{11, 12, 15}.}
\\
\vspace{.5 cm}
\begin{tabular}{|c|c|c|c|c|c|c|}
\hline
WD & Alt ID & $M_{0}\pm\triangle M$ & $R_{0}\pm\triangle R$ &$T_{eff}\pm\triangle T_{eff} (K)$ & $\alpha$ & $ R_{*} (m)$ \\
\hline
0642-166 & Sirius B & 1.02$\pm$0.02 & 0.0081$\pm$0.0002 & 25193$\pm$37 & 0.975 & 278.5 \\
                                                          &&&&& 1.113 & 592 \\
\hline
0416-594 & $\varepsilon$ Ret B & 0.62$\pm$0.022 & 0.0129$\pm$0.0003 &15310$\pm$350 & 0.917 & 1178.8 \\
                                                          &&&&& 1.089 & 1366.8 \\
\hline
1105-048 & LP 672-1 & 0.45$\pm$0.094 & 0.0133$\pm$0.0026 & 15141$\pm$88 & 0.530 & 548.7 \\
                                                          &&&&& 1.380 & 347.4 \\
\hline
1143+321 & G148-7 & 0.71$\pm$0.072 & 0.0149$\pm$0.0010 & 14938$\pm$96 & 0.768 & 1124.5 \\
                                                          &&&&& 1.255 & 1845.2 \\
\hline
1327-083 & W485 & 0.53$\pm$0.079 & 0.0141$\pm$0.00085 &13920$\pm$167 & 0.846 & 17338 \\
                                                          &&&&& 1.305 & 2489.3 \\
\hline
2341+322 & LP 347-6 & 0.56$\pm$0.022 & 0.0124$\pm$0.0007 &12300$\pm$148 & 0.790 & 1039.6 \\
                                                                &&&&& 1.230 & 1573.6 \\
\hline
\end{tabular}
\end{center}
\vspace{.7 cm}

\noindent
\vskip 1.5 cm

In conclusion, we have considered the UG hydrostatic equilibrium equation in the framework of polytropic and degenerate gas models for selected WDs, from which we obtain bounds on the characteristic length, $R_{*}$, and scaling dimension, $\alpha$, of the UG model. For ultra-massive WDs, in order to get reliable bounds on the UG parameters, one may include general relativity corrections in UG hydrostatic equilibrium equation. The effect of UG shows that WDs heavier than the Chandrasekhar mass limit might exist. The location of WDs in the H-R diagram is also shown to be affected by UG.

\pagebreak

\end{document}